\documentstyle[12pt]{article}
\def\bfm#1{\mbox{\boldmath$#1$}}
\newcommand{\dsp}{\displaystyle}
\begin{document}

\begin{titlepage}

{\noindent
{\Large \bf
LA-UR-97-3787, LANL, Los Alamos (1997);\\
to be published in {\em Nucl. Instr. Meth. A}\\
}}

\vspace*{1cm}

\begin{center}

{\Large \bf

Experimental and computer simulation study of the radionuclides
produced in thin
$^{209}$Bi targets by 130 MeV and 1.5 GeV protons
\\ }

\vspace*{0.5cm}

{
\bf
Yu.E. Titarenko $^{a,}$\renewcommand{\thefootnote}{ *}\footnote{Corresponding
author. Tel.: +7 095 125\,9100, fax: +7 095 127\,0543, e-mail:
titaren@vitep5.itep.ru},
O.V. Shvedov$^{a}$, M.M. Igumnov$^{a}$,
R. Michel$^{b}$,\\
S.G. Mashnik$^{c,f}$, E.I. Karpikhin$^{a}$, V.D. Kazaritsky$^{a}$,
V.F. Batyaev$^{a}$,\\
 A.B. Koldobsky$^{d}$, V.M. Zhivun$^{d}$, A.N. Sosnin$^{e}$,
R.E. Prael$^{f}$,\\ M.B. Chadwick$^{f}$, T.A.Gabriel$^{g}$,
M. Blann$^{h}$
}\\

\vspace*{0.5cm}

\begin{footnotesize}
$^a${\em SSC Institute for Theoretical and Experimental Physics,
B. Cheremushkinskaya 25, Moscow, 117259, Russia }\\
$^{b}~${\em Center for Radiation Protection and Radioecology,
University Hannover, Am Kleinen Felde 30, D-30 167 Hannover, Germany}\\
$^{c}~${\em Bogoliubov Laboratory of Theoretical Physics, JINR,
Dubna, Moscow Region, 141980, Russia}\\
$^{d}~${\em Moscow Engineering Physical Institute,
Moscow, 115409, Russia}\\
$^{e}~${\em Laboratory of Computing Technique and Automation, JINR,
Dubna, Moscow Region, 141980, Russia}\\
$^{f}~${\em Los Alamos National Laboratory,  Los Alamos, NM 87545, USA}\\
$^{g}~${\em Oak Ridge National Laboratory, Oak Ridge, TN 37831, USA}\\
$^{h}~${\em 7210E Calabria ct, San Diego,  CA 92122, USA}
\end{footnotesize}

\end{center}

\vspace*{0.8cm}

\begin{abstract}
The results of experimental and computer simulation studies of
the yields of residual product nuclei in $^{209}$Bi thin targets
irradiated
by 130 MeV and 1.5 GeV protons are presented. The yields  were
measured by direct high-precision $\gamma$-spectrometry.
The $\gamma$-spectrometer resolution was 1.8 keV in the 1332 keV line.
The $\gamma$-spectra were processed by the ASPRO code.
The $\gamma$-lines were identified, and the cross sections defined,
by the SIGMA code using the GDISP
radioactive database. The process was monitored by the
$^{27}$Al(p,x)$^{24}$Na reaction. Results are presented for
comparisons between the $^{209}$Bi(p,x) reaction yields obtained
experimentally and simulated by the HETC, GNASH, LAHET,  INUCL, CEM95,
CASCADE, and ALICE codes.
\end{abstract}

\end{titlepage}

\newpage

\section*{1. Introduction }

Ecology has become a matter of priority in all of the developed countries.
Utilization and management of wastes from nuclear power plants
is a priority
among the environmental protection measures. Besides, the nations that
posses nuclear weapons have become concerned with disposal or peaceful uses
(conversion) of surplus weapon plutonium and of highly-enriched uranium.
Appropriate approaches are under development in Europe, North America,
Japan, and
Russia~\cite{Rubia,Bowman,Takizuk,Kazar}.

In practice, two basic approaches are under study, namely, the safe
long-term storage of long-lived radioactive wastes, which postpones
the need for
destroying the wastes, and nuclear transmutation, which turns the wastes
into stable and shorter-lived nuclides. Some methods that combine the
two approaches are also being developed.

The concept of accelerator-driven electronuclear facilities was suggested some
four
decades ago \cite{Vas} and developed persistently ever since
is considered to be a
promising technology for waste transmutation. Such facilities
differ from conventional
nuclear
power facilities that
are operated in critical mode supported by delayed neutrons. An accelerator
provides the neutrons to sustain a chain fission reaction, so that such
facilities are subcritical. The neutron source is a nucleon-meson
cascade initiated by 0.8-2.0 GeV protons that strike a target located
inside the facility. In terms of neutron yield, nonfissible Pb, Bi, W, and
Hg, as well as fissible U and Th, are most suitable as target
materials. Composite targets with a suppliment of
the long-lived $^{99}$Tc and $^{90}$Sr
fission products may also be interesting.

The neutron-generating target provides a neutron flux that is sufficient
for effective transmutation to occur in a neutron-multiplying blanket that
contains nuclear fuel and long-lived nuclear wastes.

The accelerator-driven facilities are advantageous mainly because
of their subcritical mode of operation,
thus precluding any emergencies accompanied by an
uncontrollable rise of power. This is primarily due to the possibility for
proton accelerators to be controlled, or else de-energized promptly, so
that the power of the facility blanket falls completely.

Current
development of neutron-generating targets for accelerator-driven nuclear
facilities is aimed at extending the data on the numerous reactions induced
by high-energy neutrons and protons in target materials, including
generation of product nuclei within the range of atomic numbers A from 1 up to
 the target atomic number plus 1. The generated nuclei may be both
radioactive
and
stable. Their nuclear characteristics may affect the rated performances and
safe operation of a facility through:
\begin{itemize}
\item total activity of the target,
\item ``poisoning" the target,
\item accumulation of long-lived radionuclides to be transmuted,
\item alpha-activity of the isotopes (e.g., Po) produced,
\item production of nuclides with a high vapor pressure (T, He, Hg, etc.),
\item accumulation of chemically-active nuclides,
thus reducing the corrosion resistance of the structure materials.
\end{itemize}

The present-day accuracy requirements of predicting the yields of
product nuclei are estimated to be about 30\% \cite{Konin}. The major
difficulty
is that experiments with heavy target nuclei can only detect from 80 to 100
product nuclei, whereas the actual nuclide production yields are an order
of magnitude higher. It has been estimated that only 300-400 of the long-lived
nuclides are of interest, so we have to use mainly the
theoretically-predicted reaction cross sections.
The predictive power of the
computational methods and of the respective computer simulation codes can
be improved by making comparisons with the still scanty available
experimental data \cite{Mich1,Mich2}. In his review of nuclide yield data,
A.J. Koning \cite{Konin} has shown that the NSR and EXFOR databases are
insufficient at
energies above 20 MeV.

With the above approach, the accuracy requirements may be met by step
by step improvements in the computer simulation of intranuclear processes, as
well as by extending the experimental data base, in particular the data on
the nuclide yields in thin targets irradiated by up to 2 GeV protons. A
comprehensive comparison between the codes and the experimental data is
expected to help determine the ranges of energies and mass numbers described
most adequately by the codes. In the given work the energy choice is caused
by the presence in a target proton
spectrum of two basic components which actively participate in secondary
nuclides
formation: the primary one, with energy 1-2 GeV, created directly due to
protons in the
accelerator beam; and a secondary one, with energy
basically up to $\sim$ 0.2 GeV,
formed in primary proton interactions  with target nucleus.

\section*{2. Techniques}
Generation of residual product nuclei in proton interactions with a
target nucleus is due to the intranuclear processes of spallation, fission,
fragmentation, and evaporation of light nuclei and nucleons.
The general form of proton interactions with a target nucleus is

\begin{equation}
{}^{A_T}_{Z_T}T(p,x)^A_ZN \mbox{ ,}
\label{a1}
\end{equation}
where, as usual,
$(p,x)$ designates a nuclear reaction type,
$T$ and $N$ are chemical symbols of the
elements, i.e. of the target nucleus and
of product nuclei,
$A_T$ and $Z_T$ are respectively the mass number and the charge of a target
nuclide,
$A$ and $T$ are respectively the mass number
and the charge of a nuclide produced
in the respective nuclear reaction.

The present work determines the independent and cumulative yields of
radioactive nuclear reaction products. The $^{27}Al(p,x)^{24}Na$
reaction is used as monitor.

The variations in the concentrations of two ``congener" nuclides produced
in
a thin target under irradiation may be presented as
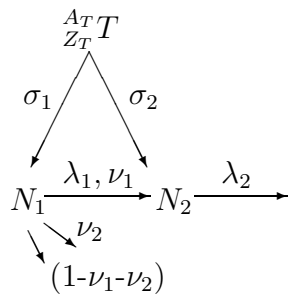
\begin{center}
\begin{picture}(130,100)
\put(40,95){${}^{A_T}_{Z_T}T$}
\put(50,90){\vector(-1,-2){22}}
\put(50,90){\vector(1,-2){22}}
\put(20,30){$N_1$}
\put(75,30){$N_2$}
\put(33,35){\vector(1,0){40}}
\put(25,70){$\sigma_1$}
\put(65,70){$\sigma_2$}
\put(40,40){$\lambda_1, \nu_1$}
\put(90,35){\vector(1,0){35}}
\put(100,40){$\lambda_2$}
\put(33,25){\vector(4,-3){12}}
\put(27,22){\vector(1,-2){6}}
\put(45,20){$\nu_2$}
\put(35,1){(1-$\nu_1$-$\nu_2$)}
\end{picture}
\end{center}
\begin{equation}
\label{2}                                                              
\cases{
\frac{\dsp dN_1}{\dsp dt}=\sigma_1N_T\Phi-\lambda_1N_1 \mbox{ ,}
\cr\cr
\frac{\dsp dN_2}{\dsp dt}=\sigma_2N_T\Phi+\nu_1\lambda_1N_1-
\lambda_2N_2 \mbox{ ,}\cr}
\end{equation}

{\noindent
with the initial conditions $N_1(0)=0$ and $N_2(0)=0$. Here,
$N_1$ and $N_2$ are the numbers of nuclei produced,
$\sigma_1$ and $\sigma_2$ are their yields,
$\lambda_1$ and $\lambda_2$ are the decay constants,
$\nu_1$ and $\nu_2$ are the probabilities for nuclide 1 to decay into
nuclide 2 and 3, i.e. the branching factors,
$N_T$ is the number of $^{A_M}_{Z_M}T$ in an experimental sample,
$\Phi$ is the proton flux and $t$ is the current time.
}
In the case of a pulsed irradiation mode, then, the solution for the set of
differential equations (\ref{2}) is described by the expressions

\begin{eqnarray}
\label{3}                                                              
N_1\left[(K-1)T+\tau\right] &=& \frac{N_T\Phi\sigma_1F_1}{\lambda_1}
\mbox{ ,}\\
N_2\left[(K-1)T+\tau\right] &=& N_T\Phi\nu_1\frac{1}{\lambda_2-\lambda_1}
\sigma_1F_1+
\frac{N_T\Phi}{\lambda_2}\left[\sigma_2-
\frac{\lambda_1}{\lambda_2-\lambda_1}\nu_1\sigma_1\right]F_2 \mbox{ ,}  
\end{eqnarray}

{\noindent
where
$\tau$ is the duration of a single proton pulse,
$T$ is the pulse repetition period,
$K$ is the number of irradiation pulses, and
}

$$F_i=\left( 1-e^{-\lambda_i\tau} \right) \frac{1-e^{-\lambda_iKT}}
{1-e^{-\lambda_iT}}\mbox{ ,} \qquad i=1, 2, Na\mbox{ .}$$

The after-irradiation decay of the nuclides produced is described by
the set

\begin{equation}
\label{5}                                                             
\cases{
\frac{\dsp dN_1}{\dsp dt}=-\lambda_1N_1 \mbox{ ,}\cr\cr
\frac{\dsp dN_2}{\dsp dt}=\nu_1\lambda_1N_1-\lambda_2N_2 \mbox{ .}\cr}
\end{equation}

The solution for (\ref{5}) is

\begin{eqnarray}
N_1(t) &=& N_{1_0}e^{-\lambda_1t} \mbox{ ,} \\                         
N_2(t) &=& \left[N_{2_0}+\frac{\lambda_1}{\lambda_1-\lambda_2}
\nu_1N_{1_0} \right] e^{-\lambda_2t} - \frac{\lambda_1}{\lambda_1-
\lambda_2}
N_{1_0}\nu_1e^{-\lambda_1t} \mbox{ ,}                                  
\end{eqnarray}

{\noindent
where
$N_{1_0}$ and $N_{2_0}$ are the
numbers of nuclei produced as the ``cooling" starts
(i.e., the
irradiation stops).
}

The number of nuclei produced as the irradiation stops corresponds to their
number as the ``cooling" starts. In this case the following condition is
satisfied for $K$ irradiation pulses:

\begin{equation}
\label{8}                                                               
N_{1_0}=N_1\left[(K-1)T+\tau\right] ~,~
N_{2_0}=N_2\left[(K-1)T+\tau\right] \mbox{ .}
\end{equation}

Instead of the numbers of nuclei in experimental samples, i.e., the nuclear
concentration, the actual experiment measures the counting rates in the
total absorption peaks at  $\gamma$-line energies $E_1$ and $E_2$, i.e.,
the intensities in the peaks that are related to the concentrations
$N_1(t)$ and $N_2(t)$ as

\begin{equation}                                                       
S_1(t)=N_1(t)\lambda_1\eta_1\varepsilon_1 ~,~
S_2(t)=N_2(t)\lambda_2\eta_2\varepsilon_2  \mbox{ ,}
\end{equation}

{\noindent
where
$\eta_1$ and $\eta_2$ are the yields of the $\gamma$-lines and
$\varepsilon_1$ and $\varepsilon_2$ are the spectrometric effectiveness at
$\gamma$-line energies $ E_1 $ and $ E_2$.
}
Applying, then the condition (\ref{8}) to $S_1(t)$ and $S_2(t)$ ), we get

\begin{eqnarray}
\label{10}                                                             
S_1(t) &=& A_0e^{-\lambda_1t} \mbox{ ,} \\
\label{11}                                                             
S_2(t) &=& A_1e^{-\lambda_1t}+A_2e^{-\lambda_2t} \mbox{ ,}
\end{eqnarray}

{\noindent
where
}

\begin{eqnarray}
\label{12}                                                          
A_0&=&N_T\Phi\sigma_1\eta_1\varepsilon_1F_1 \mbox{ ,} \\
\label{13}                                                          
A_1&=&N_T\Phi\sigma_1\eta_2\varepsilon_2
\frac{\lambda_2}{\lambda_2-\lambda_1}\nu_1F_1 \mbox{ ,} \\
\label{14}                                                           
A_2&=&N_T\Phi\eta_2\varepsilon_2 \left(\sigma_2-\sigma_1\nu_1
\frac{\lambda_1}{\lambda_2-\lambda_1} \right) F_2 \mbox{ .}
\end{eqnarray}

{\noindent
The coefficients $A_0$, $A_1$ and $A_2$ that carry information on the
cross sections $\sigma_1$ and $\sigma_2$  are determined by least
squares-fitting the experimental points $S_{1i}$ and $S_{2i}$ through the
respective functions
}

\begin{eqnarray}
\label{15}                                                          
g(t)&=&A_0e^{-\lambda_1t} \mbox{ ,} \\
\label{16}                                                          
f(t)&=&A_1e^{-\lambda_1t}+A_2e^{-\lambda_2t} \mbox{ .}
\end{eqnarray}

One constructs the quadratic functionals $R_1$ and $R_2$
\begin{eqnarray}
R_1&=&\sum_{i=1}^{L_1}{\left(S_{1i}-A_0e^{-\lambda_1t_i}\right)}^2
/\Delta S^2_{1i} \mbox{ ,} \\                                          
R_2&=&\sum_{i=1}^{L_2}{\left(S_{2i}-A_1e^{-\lambda_1t_i}-
A_2e^{-\lambda_2t_i}\right)}^2/\Delta S^2_{1i} \mbox{ ,}               
\end{eqnarray}

{\noindent
where
$S_{1i}$ and  $S_{2i}$ are the counting rates measured in the total absorption
peaks of $\gamma$-quanta of the first and second nuclide at moment $t_i$,
$\Delta S_{1i}$ and $\Delta S_{2i}$ are the absolute errors in the above
counting rates and
$t_i$ is the time span from the irradiation end to the middle of the $i$-th
measurement interval.
}
If
$L_1$ and  $L_2$ are the numbers of
experimental points for the first and second
nuclide, respectively, then
using the condition of minimizing the functionals, we get the following
expressions to find the parameters $A_0$, $A_1$, $A_2$ and their errors:

\begin{eqnarray}
\label{19}
A_0&=&\frac{\dsp \sum_{i=1}^{L_1}
\left( S_{1i}e^{-\lambda_it_i}/\Delta S^2_{1i} \right)}
{\dsp \sum_{i=1}^{L_1}
\left( e^{-2\lambda_it_i}/\Delta S^2_{1i} \right)} \mbox{ ,} \\         
\Delta A_0&=&\left(\sum_{i=1}^{L_1}
\frac{e^{-2\lambda_it_i}}{\Delta S^2_{1i}}\right)^{-1/2} \mbox{ ,}      
\label{20}
\end{eqnarray}
\begin{eqnarray}
\label{21}
\overrightarrow{A}&=&M^{-1}\overrightarrow{Z} \mbox{ ,} \\           
\Delta A_i&=&\sqrt{\left(M^{-1}\right)_{ii}}\mbox{ ,}
\qquad (i=1,2)\mbox{ ,}                                              
\label{22}
\end{eqnarray}

{\noindent
where}

$$\overrightarrow{A}=\left\{A_1,A_2\right\}, \qquad
 \overrightarrow{Z}=\left\{Z_1\atop Z_2\right\} \mbox{ .} $$

{\noindent
The matrix $M$ and the vector of the right-hand side
$\overrightarrow{Z}$ of the initial set of linear equations are
}

\begin{eqnarray}
\label{23}
M_{ij} &=& \sum_{k=1}^{L_2}\left(e^{-(\lambda_i+\lambda_j)t_k}/
\Delta S^2_{2k}\right) \mbox{ ,} \\                                   
Z_i &=& \sum_{k=1}^{L_2}\left(S_{2k}e^{-\lambda_it_k}/
\Delta S^2_{2k}\right) \mbox{ ,}  \qquad i,j=1,2  \mbox{ .}           
\label{24}
\end{eqnarray}

Calculating the cross sections necessitates determination of the proton flux
$\Phi$. With that purpose, an experimental sample was irradiated together
with the Al monitor sample for which we have, by analogy with expressions
(\ref{3}), (\ref{10}) and (\ref{12})

\begin{equation}
\label{25}
S_{Na}(t)=N_{Al}\Phi\sigma_{st}F_{Na}\eta_{Na}\varepsilon_{Na}
e^{-\lambda_{Na}t}=Be^{-\lambda_{Na}t} \mbox{ ,}                      
\end{equation}

{\noindent
where $\sigma_{st}$ is $~^{27}Al(p,x)^{24}Na$ monitor reaction cross section.
}
The parameter $B$ is also determined by least squares fitting the
experimental points through a dependence of the form (\ref{15}) using
formulas (\ref{19}) and (\ref{20}).
The number of ${}^{24}$Na nuclei produced in the monitor will, then, be

$$N_{Na}=\frac{B}{\eta_{Na}\varepsilon_{Na}\lambda_{Na}}=
N_{Al}\Phi\sigma_{st}\frac{F_{Na}}{\lambda_{Na}} \mbox{ ,} $$

{\noindent
which permits the proton flux $\Phi$ to be presented as
}

\begin{equation}
\label{26}
\Phi=\frac{N_{Na}\lambda_{Na}}{N_{Al}\sigma_{st}F_{Na}} \mbox{ .}      
\end{equation}

The expressions (\ref{12}), (\ref{19}), and (\ref{26}) can be used to
calculate the
cumulative yield of the first nuclide (or its independent yield in case its
precursors are absent):

\begin{equation}
\label{27}
\sigma_1^{cum}=\frac{A_0}{\eta_1\varepsilon_1F_1N_{Na}}
\frac{N_{Al}}{N_T}\frac{F_{Na}}{\lambda_{Na}}                         
\sigma_{st}  \mbox{ .}
\end{equation}

{\noindent
At the same time, the formulas (\ref{13}), (\ref{14}), (\ref{21}) and
(\ref{26}) can be used to
obtain expressions for calculating the cumulative yield of the first
nuclide, as well as the independent and cumulative yields of the second
nuclide:
}

\begin{eqnarray}
\label{28}
\sigma_1^{cum} &=& \frac{A_1}{\nu_1\eta_2\varepsilon_2F_1N_{Na}}
\frac{N_{Al}}{N_T}\frac{\lambda_2-\lambda_1}{\lambda_2}
\frac{F_{Na}}{\lambda_{Na}}\sigma_{st}  \mbox{ ,} \\                   
\label{29}
\sigma_2^{ind} &=& \left(\frac{A_2}{F_2}+\frac{A_1}{F_1}
\frac{\lambda_1}{\lambda_2}\right)
\frac{1}{\eta_2\varepsilon_2N_{Na}}
\frac{N_{Al}}{N_T}\frac{F_{Na}}{\lambda_{Na}}
\sigma_{st}  \mbox{ ,} \\                                             
\label{30}
\sigma_2^{cum} &=& \sigma_2^{ind}+\nu_1\sigma_1^{cum}=
\left( \frac{A_1}{F_1}+\frac{A_2}{F_2} \right)
\frac{1}{\eta_2\varepsilon_2N_{Na}}
\frac{N_{Al}}{N_T}\frac{F_{Na}}{\lambda_{Na}}
\sigma_{st}  \mbox{ .}                                                
\end{eqnarray}

{\noindent
Obviously, the yields calculated by formulas (\ref{27}) and (\ref{28}) must
be the
same. However, the yield obtained by formula (\ref{27}) is usually included
in the
final results because the calculation accuracy of (\ref{27}) is higher.
}

\section*{3. Monitor reactions}

As noted above, the
$^{27}$Al(p,x)$^{24}$Na reaction was used as a monitor. The
$^{27}$Al(p,x)$^{24}$Na cross section was
calculated from the approximation function
\begin{equation}
\label{mon}
\sigma=\sum_{i=0}^ka_iE^i
\end{equation}
recommended by V.G. Khlopin Radium Institute with the coefficients
tabulated in Table~1~\cite{Khlop}.

It should be noted  that Cumming \cite{Cum} was the first
(in 1963) who had obtained and recommended to use the values of the
$^{27}$Al(p,x)$^{24}$Na reaction excitation function. At present, however,
the use of this reaction to monitor a proton flux is regarded as incorrect
because the (n,$\alpha$) reaction may also contribute to the yield of
$^{24}$Na, thus
overestimating the proton flux density $\Phi$. Therefore, the
$^{27}$Al(p,x)$^{22}$Na monitor has been preferred recently. The
situation  can be seen clearly in Fig. 1 showing the excitation functions
of the two reactions evaluated independently at the V.G. Khlopin Radium
Institute and used at ZSR (Germany) \cite{Mich3}.
A comparison of these two plots corroborates the advantage of using the
$^{27}$Al(p,x)$^{22}$Na reaction as a monitor.
At the same time, this monitor does not
seem to be very attractive for us because we  have to measure the
short-lived (p,x) reaction products, which is particularly urgent in studying
medium and heavy nuclei with reaction products form complicated decay
chains. This condition restricts the irradiation times of the experimental
samples  since only long-lived nuclides are accumulated due to larger amont of
irradiation times and hence the loading parameters of the
spectometer are becoming
worse.
For the minor accumulation of
$^{22}$Na resultant from the selected short irradiation times to be
measured within a high accuracy, we have to use low-background
spectrometers, thus making the research much more expensive.

One can see from Fig. 1(A) that the data used at ZSR are above
the V.G.Khlopin Radium Institute recommended curve by $\sim$ 10\% at 100-200
MeV, by $\sim$ 5\% at 600-1600 MeV and by $\sim$ 12.5\% at 1600-2600
MeV. Since  the accuracy of the two data groups is 10\%, they can be
regarded as coincident to within errors. Therefore, the possible neutron
background becomes very important because its occurrence during irradiation
of experimental samples would additionally increase the systematic
differences among the eventual values of yields obtained by different
researchers with different monitor reactions.

\section*{4. Irradiation of Experimental Samples}

The yields of residual product nuclei from 1500 MeV proton
irradiation were determined by exposing the experimental samples to the
beam
extracted by slow extraction system from the ITEP U-10 synchrotron. The
extraction is schematically presented in Fig. 2.
The extracted beam has the form of an ellipse  with $\sim$ 25x15 mm axes, the
beam
intensity is  $\sim 2\cdot 10^{11}$ proton/pulse, the pulse repetition rate
$\sim$ 16 min$^{-1}$, duration of a single pulse $\sim$ 0.5 s.

The yields of residual product nuclei from 130 MeV proton irradiation
were determined by exposing the experimental samples to the proton beam used
for medical purposes. The extraction is schematically presented in Fig. 3.
The beam was formed by a set of collimators. The beam at the outlet of
the last collimator is of circular form of $\sim$ 15 mm diameter,
$\sim 5 \cdot 10^{9}$ proton/pulse intensity, $\sim$ 16 min$^{-1}$ pulse
repetition rate, and $\sim$ 100 ns single pulse duration.

Use was made of experimental samples of metallic Bi prepared by pressing
metallic powder. Before the exposure, the experimental samples were weighed
with Sartorius BP-61 analytical scales and were then ``soldered" together
with
the monitor samples and interlayers into polyethylene envelopes.

In each of the irradiation runs, a monitor-Al interlayer-experimental
sample sandwich stack was placed normally to the proton beam of the
respective energy. The diameters of experimental sample, interlayer and
monitor were strictly the same (10.5 mm). The exposure time of a single
sample was  $\sim$15-30 min. The 1500 MeV and 130 MeV proton fluences were
$\sim 2.4 \cdot 10^{13}$ and  $\sim 1.5 \cdot 10^{11}$, respectively.

After the exposure, the experimental samples and the monitors were
replaced into sealed polyethylene envelopes to prevent the gaseous reaction
products from being lost.

The irradiation envelopes were used to check  the loss of the reaction
products which may escape from the irradiated experimental samples by
analyzing the composition of the samples. This check demonstrated the
complete absence of loss.

\section*{5. Estimation of the neutron background}

In the present work, the possible neutron background in exposing
experimental samples to proton beams was estimated by the SSNTD techniques
used to find the parameters of medical proton beams \cite{Lom}. With
SSNTDs, we measured the secondary radiation background composed of neutrons
and protons. In the experiment, we used the $^{209}$Bi,$^{238}$U and
$^{237}$Np targets of different fission thresholds and the glass plate
detectors with collimator grids. The detectors and the targets were placed
near the irradiated samples at different distances from the proton beam. In
calibration, the detectors and the targets were irradiated by a proton beam
for a short time. Fig. 4 shows the relative distributions of the track
numbers from the fission fragments recorded with the SSNTDs near the
extracted 1500 MeV and 130 MeV proton beams.

The results displayed in Fig. 4 show that the secondary background is
nearly isotropic and that the total number of fission fragments due to the
secondary radiation near the two beams is as small as a few hundredths of
one per cent of the number of the recorded fission fragments in the proton
beams proper. This indicates that such a background  cannot introduce any
substantial distortions to the proton flux density found using the
$^{27}$Al(p,x)$^{24}$Na reaction excitation function. Surely, this is valid
only if the data selected to plot the recommended dependence do not
comprise the neutron background error.

\section*{6. Measurement and processing of $\gamma$-ray spectra}

The measurement facility is a spectrometric circuit comprising a GC-2518
Ge detector, a 1510 module (ADC, amplifier, high-voltage supply), and a
S-100 base that as an integral part of an IBM PC emulates a multichannel
analyzer. The $^{60}$Co 1332 keV $\gamma$-line energy resolution of the
facility
is 1.8. keV. Fig. 5 presents the measured $\gamma$-spectra of the $(p,x)$
reaction products.

The $\gamma$-spectra measured were processed by the PC IBM-realized ASPRO
code \cite{Atr}. The spectra thus processed were combined to form an input
file
for the SIGMA code. The code plots the intensity of a selected
$\gamma$-line
versus time and, using its energy and simulated half-life, identifies the
produced nuclides by the GDISP  database and simulates their cross sections
by formulas (\ref{27}-\ref{30}) \cite{Spa}.

The measured $\gamma$-spectra are of high intensity, especially during the
early decay period. Therefore, any possible detection losses under high
loads
of the spectrometer must be checked out carefully. To avoid the associated
systematic errors, the spectrometer was checked out by the method of two
sources. Fig. 6 shows the variations in the peak area and resolution
versus the spectrometer load.

Conforming to these results, the ultimate load of the
spectrometer was limited so as not to exceed 5\% in all of the measurements.
To remain within that limitation, we started monitoring the experimental
samples at
$\sim$500 mm height and, as the load decreased, descended down to the
ultimate
height of 40 mm. The $\gamma$-spectra measured were correlated to each
other
by introducing the height factors determined by the function
$R(x)=a/(b+x)^2$
with constants
calculated by approximating
the experimental points. Fig. 7 shows a plot of this function.

The independent and cumulative yields of the reaction products in formulas
(\ref{27}-\ref{30}) were calculated using the relative spectrometric
effectiveness. Fig.
5 shows the analytical dependence of the spectrometric effectiveness in the
36.5-2650.0 keV range.
The techniques of determining the effectiveness are described in
\cite{Np_Th}.

In addition, we have used in our measurements also
a low-background spectrometer with  $\sim$50
and $\sim$100  decay/hour ultimate sensitivities of its Ge-Li detector at
1500
keV and 200-300 keV, respectively. This spectrometer was designed for
measuring the magnetic moment of the
neutrino and has a counterpart described in
\cite{Star}. This spectrometer was used to measure $\gamma$-spectra of
experimental samples 1.5 years after the samples were irradiated.
Cross sections of long-lived nuclear reaction products were obtained
by identifying and analyzing these $\gamma$-spectra.

The corrections for $\gamma$-absorption in the substances of
experimental or monitor samples were calculated using a expression
published in \cite{Zwei}. The values of the total linear attenuation factor
$\mu$ were taken from \cite{Storm}.

\section*{7. Experimental Results }

Tables 2 through 5 present the results of measuring the independent and
cumulative yields of the ground and metastable states of the $^{209}$Bi
nuclear reaction products at proton energies of 1500 MeV and 130 MeV.
Table 2
presents the experimental cumulative yields of the ground states of the
reaction products, Table 3, the experimental and simulated
independent yields of the ground states of the reaction products, Table 4,
the experimental and simulated independent and cumulative yields
of
the metastable states of the reaction products and Table 5,
the total
experimental yields of the ground and metastable states together with the
simulated yields of the ground states of the reaction products.

The $^{27}$Al(p,x)$^{24}$Na cross sections were calculated according to
(\ref{mon}) and are
taken to be
9.9$\pm$1.0 mb and 9.6$\pm$1.0 mb at 1500 MeV and 130 MeV proton energies,
respectively.

It should be noted that the experimental values have been renormalized
to the quantum yields in the PCNUDAT nuclear
database \cite{Kins} and Table of Isotopes (8th Edition) \cite{Fire}.
The renormalization has become necessary because the
tabulated experimental yields of reaction products were obtained for many
of
the nuclides by averaging over several $\gamma$-lines. When varying the
recommended $\gamma$-quantum yield data, therefore, their separate values
have to be
renormalized and averaged again. The complete information cannot be
published
in practice because it is too voluminous.
The reaction yields obtained using the low-background spectrometer are
labelled (*).

Fig. 8 shows some of the most characteristic plots of the time dependences
of the $^{209}$Bi(p,x) reaction product decays.

In the case of  $^{209}$Bi irradiated by 1500 MeV protons, Fig. 8(A) shows
the time dependencies of two independent $^{179}$Re and $^{198}$Pb decays
at E$_\gamma$$\sim$290.0 keV and of the independent $^{95}$Tc decay
together with the decay chain $^{95}$Zr$\to$$^{95}$Nb of parent nuclides
at E$\gamma$$\sim$765.8 keV. In the case of  $^{209}$Bi irradiated by 130
MeV protons, Fig. 8(B) shows the time dependencies of the independent
$^{205}$Po decay at E$\gamma$=872.4 keV, of the decay chain
$^{201}$Bi$\to$$^{201}$Pb of parent nuclides at E$\gamma$=331.2 keV.
{}From
Fig. 8 it can be seen that the
analyzing the decay
curves of the nuclear reaction products may be much more complicated
compared with the
simplified case described above in the section 2.
This is due, first of all, to the fact that the nuclear
transition energies of the reaction products may prove to be so alike that the
actual unresolvable $\gamma$-lines occur in the measured $\gamma$-spectra.
Given the situations like the above, the SIGMA code realizes the
feasibility of separating them by the least squares method and in view of
the fact that the nuclides produced are of  different half-lives. One of
the SIGMA subroutines separates the nuclide decays by independent decay
chains, while another separates an independent nuclide decay from a chain
of parent nuclides.

\section*{8. Measurement errors}

Analyzing the errors in the results obtained has shown that they fall
mostly within
$\cong (10 \div 35)\%$. The values of the errors were calculated as follows.
Since the results presented were obtained mostly by averaging several
$(\sigma_i\pm\Delta \sigma_i)$ values calculated for different
$\gamma$-lines, their means and their errors were calculated
using formulas

\begin{equation}
\label{32}
\overline{\sigma}=\frac{\sum_i\sigma_iW_i}{\sum_iW_i}                  
, \qquad \mbox{where} \qquad W_i=1/\Delta \sigma_i^2 \mbox{ ,}
\end{equation}

\begin{equation}
\label{er1}
\Delta \overline{\overline{\sigma}}=
\sqrt{\frac{\sum_iW_i\left(\overline{\sigma}-\sigma_i\right)^2}
{\left(n-1\right)\sum_iW_i}} \mbox{ .}                                
\end{equation}

The total error in the values presented in Tables 2, 3, 4, 5 was calculated
making allowance for the monitor error as

\begin{equation}
\label{er2}
\frac{\Delta \overline{\sigma}}{\overline{\sigma}}=
\sqrt{\left(\frac{\Delta \overline{\overline{\sigma}}}
{\overline{\sigma}} \right)^2 +
\left( \frac{\Delta \sigma_{st}}{\sigma_{st}}\right)^2} \mbox{ .}    
\end{equation}

The errors in the independent and cumulative yields of the reaction
products for separate $\gamma$--lines, $\Delta \sigma_i$ calculated by
formulas (\ref{er1}) and (\ref{er2}) were determined using the formulas
for transfer of errors.

As shown by the analysis, the uncertainties in the nuclear data on
$\gamma$-yields and on monitor reaction cross sections make the major
contribution to the total error.

\section*{9. Theoretical simulation of measured products}

\subsection*{9.1. Technique of comparing experimental and simulated results}

The simulation results should be compared with the experimental data making
allowance for the fact that any of the computer simulation codes can only
simulate the independent yields of nuclear reaction products, whereas most
of the experimental data are the cumulative yields. So, any correct
comparison must involve a procedure for calculating a simulated cumulative
yield on the basis of the simulated independent yields of the respective
precursors.

If the production chain of n nuclides is presented as

$$\begin{array}{ccccccc}
\downarrow\lefteqn{\sigma_1}&&\downarrow\lefteqn{\sigma_2}&&
\ldots&&\downarrow\lefteqn{\sigma_n}\\
\fbox{\mathstrut 1}&\stackrel{\nu_1,\lambda_1}{\rightarrow}&
\fbox{\mathstrut 2}&\stackrel{\nu_2,\lambda_2}{\rightarrow}&
\ldots&\stackrel{\nu_{n-1},\lambda_{n-1}}{\rightarrow}&
\fbox{\mathstrut n}
\end{array} \mbox{ ,}$$

{\noindent
the cumulative yield of the last (n-th) nuclide can be calculated as
}
\begin{equation}
\label{35}
\sigma_n^{cum}=\sum_{i=1}^n\sigma_i^{ind}l_i \mbox{ ,}                
\end{equation}

{\noindent
where $l_i$ is the fraction of decays of nuclide $i$ that go to nuclide
$n$, which is defined using the branching factors $\nu_j$ as
}
$$\cases{
\begin{array}{ll}
l_i=\prod\limits_{j=1}^{n-1}\nu_j,  & i<n \mbox{ ,}\cr\cr
\\
l_i=1, & i=n \mbox{ .}\cr
\end{array}
}
$$

The above representation ensures from the conceptual physical treatment
of the independent and cumulative yields and is corroborated by the form of
formulas (\ref{28}-\ref{30}). The values of the nuclear chain branching
factors have been
taken from \cite{Gus} . Unfortunately, lack of any fresher systematized
data has made us use the somewhat obsolete values of the factors. The data
are very difficult to extract from the renewed databases, and the
extraction procedure falls outside the scope of the present research
program. Surely, use of any newer data may have affected the accuracy of
the simulated cumulative yields of the reaction products. At the same time,
identical values are very important to use when comparing the
experimental data and the simulation results obtained by different codes.

To get a correct comparison among the data simulated by different codes,
the simulation results are to be renormalized to unified the cross section
of the proton-nucleus interaction, using available experimental data
values or phenomenological formulas. Here we used Letaw's formulae
described in \cite{Pear} to obtain the unified
cross section to be 1663 mbarn at 130 MeV and 1873 mbarn at 1500 MeV.

If a comparison event with
simulation differing from experiment by not more than a factor of 2 is
assumed here to be the coincidence criterion, then the predictive
power of the codes can
be presented via a ratio of the number of coincidences to the total
number of comparison events. Another
parameter proposed in \cite{Mich2} to present  code
accuracies is the mean deviation of
simulation results from experimental data:

\begin{equation}                                                    
\label{36}
\langle H \rangle = 10^
{\sqrt{\dsp \langle\left(lg\left( \frac{\sigma_{cal}}{\sigma_{exp}}
\right)\right)^2\rangle}} \mbox{ ,}
\end{equation}

{\noindent
where $\langle~\rangle$ designates averaging over all of the comparison
events.
}

\subsection*{9.2. Computer simulation codes}
The products of $^{209}$Bi(p,x)-reaction were simulated by seven different
codes, namely,
\begin{itemize}
\item the CEM95 Cascade-Exciton Model code \cite{cem,cem95},
\item the CASCADE cascade -- evaporation -- fission transport
code \cite{cascade},
\item the INUCL cascade -- preequilibrium -- evaporation -- fission
code \cite{inucl},
\item the HETC cascade -- evaporation transport code \cite{hetc},
\item the LAHET cascade -- evaporation -- fission transport code \cite{lahet},
\item the GNASH code based on the Hauser-Feshbach and preequilibrium approach
\cite{gnash},
\item the ALICE code with HMS precompound approach
\cite{alice96}.
\end{itemize}

A detailed description of the models used by these codes
may be found in Refs.~\cite{cem}-\cite{alice96} and references given therein;
therefore, only their basic assumptions
will be outlined below. \\

CEM95 uses the Monte Carlo method to simulate hadron-nucleus
interactions in the framework of an extended version~\cite{cem95}
of the Cascade-Excitom Model (CEM) of nuclear reactions~\cite{cem}.
The CEM
assumes that the reactions occur in three stages. The first stage is
the intranuclear cascade in which primary and
secondary particles can be rescattered
several times prior to absorption by, or escape from the nucleus.
The excited residual nucleus remaining after the emission of the
cascade particles determines the particle-hole configuration that is
the starting point for the second, preequilibrium stage of the
reaction. The subsequent relaxation of the nuclear excitation is
treated in terms of the exciton model of preequilibrium decay which
includes the description of the equilibrium evaporative third stage of
the reaction.

The cascade stage of the interaction is described by the standard
version of the Dubna intranuclear cascade model (INC)~\cite{book}.
All the cascade
calculations are carried out in a three-dimensional geometry.
The nuclear matter density $\rho(r)$
is described by a Fermi distribution with two parameters
taken from the analysis of electron-nucleus scattering, namely
$$\rho(r) = \rho_p(r) + \rho_n(r) = \rho_0 \{ 1 + exp [(r-c) / a] \}
\mbox{ ,}$$
where $c = 1.07 A^{1/3}$ fm, $A$ is the mass number of the target, and
$a = 0.545$ fm.  For simplicity,
the target nucleus is divided by concentric spheres into
seven zones in which the nuclear density is considered
to be constant.
The energy spectrum of the target nucleons is
estimated in the perfect Fermi gas approximation with the local Fermi
energy
$T_F(r) = \hbar^2 [3\pi^2 \rho(r)]^{2/3}/(2m_N)$, where $m_N$ is the nucleon
mass.
The influence of intranuclear nucleons on the incoming projectile is
taken into account by adding to its laboratory kinetic
energy an effective real potential $V$, as well as by considering
the Pauli principle which forbids a number of intranuclear collisions
and effectively increases the mean free path of cascade particles inside
the target.
For incident nucleons
$V \equiv V_N (r) = T_F(r) + \epsilon$,
 where $T_F(r)$ is the corresponding Fermi
energy and $\epsilon$ is the mean binding energy of the nucleons
($\epsilon \simeq 7$ MeV~\cite{book}).
For pions, in the Dubna INC one usually
uses~\cite{book} a square-well nuclear potential
with the depth $V_{\pi} \simeq 25$ MeV, independently of the nucleus and
pion energy.
The interaction of the incident particle with the nucleus is approximated as
a series of successive quasifree collisions of the fast cascade
particles ($N$ or $\pi$) with intranuclear nucleons:
\begin{eqnarray}                                                      
\label{37}
NN \to NN , \qquad NN \to \pi NN , \qquad  NN \to \pi _1,\cdots,\pi _i NN
\mbox{ ,}
\nonumber \\
\pi N \to \pi N, \qquad  \pi N \to \pi _1,\cdots,\pi _i N  \qquad  (i \geq 2)
\ .
\label{a3}
\end{eqnarray}
To describe these elementary collisions, one uses experimental
cross sections for the free $NN$ and $\pi N$ interactions,
simulating angular and momentum distributions of secondary particles using
special
polynomial expressions with energy-dependent coefficients~\cite{book}
and one takes into account the Pauli principle.
Besides the elementary processes (37), the Dubna INC also takes into account
pion absorption on nucleon pairs
\begin{equation}                                                       
\label{38}
\pi NN \to NN .
\end{equation}
The momenta of two nucleons participating in the absorption are chosen
randomly from the Fermi distribution, and the pion energy is distributed
equally between these nucleons in the center-of-mass system
of the pion and nucleons participating in the absorption. The direction
of motion of the resultant nucleons in this system is taken as
isotropically distributed in space. The effective cross section for
absorption is related (but not equal) to the experimental cross sections
for pion absorption by deuterons.

The standard version of the Dubna INC is described in detail in the
monograph~\cite{book} and
more briefly, in the review~\cite{bar73rev} and in the recent book
by Iljinov, Kazarnovsky, and Paryev~\cite{iljinov94}.
A detailed comparison of the Dubna INC with the well known Bertini INC
developed at Oak Ridge National Laboratory \cite{bertini} and
with the popular version developed at Brookhaven National Laboratory and
Columbia University by Chen et al.~\cite{vegas} is made in
Ref.~\cite{jinrornlbnl}.

An important point of the CEM is the condition for transition from the
intranuclear cascade stage to preequilibrium  processes. In a
conventional cascade-evaporation model, cascade nucleons are traced down
to some minimal energy, the cut-off energy $T_{cut}$ being about 7--10
MeV, below which particles are considered to be absorbed by the nucleus.
In the CEM~\cite{cem}, it was proposed to relate the condition for a cascade
particle to be captured by the nucleus to the similarity of
the imaginary part of the optical potential
calculated in the cascade model $W_{opt. mod.}(r)$ to its experimental value
$W_{opt. exp.}(r)$ obtained from analysis of data on particle-nucleus elastic
scattering.
The agreement between
$W_{opt. mod.}$ and $W_{opt. exp.}$ is assumed to occur when
the proximity parameter
$${\cal P} =\mid (W_{opt. mod.}-W_{opt. exp.}) / W_{opt. exp.} \mid $$
becomes small enough.
In CEM95, a fixed value $\cal P$ = 0.3 extracted from the
analysis of experimental proton-
and pion-nucleus data at low and intermediate energies is used.

The subsequent interaction stages (preequilibrium and equilibrium)
of nuclear reactions
are considered in the CEM in the framework of an extension of the
Modified Exciton Model (MEM)~\cite{mem,modex}.
At the preequilibrium stage of a reaction the CEM takes into account all
possible nuclear transitions changing the number of excitons $n$
with
$\Delta n = +2, -2$, and 0, as well as all possible multiple subsequent
emissions of $n$, $p$, $d$, $t$, $^3$He, and  $^4$He. The corresponding
system of master equations describing the behavior of a nucleus at
the preequilibrium stage is solved by the Monte Carlo
technique~\cite{modex}.

Let us note here that in the CEM the initial configuration for the
preequilibrium decay (number of excited particles and holes, i.e.
excitons $n_0 = p_0 + h_0$, excitation energy $E^*_0$,
linear momentum ${\bfm P}_0$ and angular momentum ${\bfm L}_0$
of the nucleus) differs significantly from that
usually postulated in exciton models.

To be able to analyze reactions with heavy targets and to describe
accurately excitation functions over a wider range of incident
particle energy, the CEM has been extended recently~\cite{cem95}.
The extended version incorporates the competition between evaporation
and fission of compound nuclei, takes into account pairing energies,
considers the angular momentum of preequilibrium and evaporated
particles and the rotational energy of excited nuclei, and can use more
realistic nuclear level densities (with $Z$, $N$, and $E^*$ dependences
of the level density parameter).

The extended version of the CEM realized in the code
CEM95 is described in details in Ref.~\cite{cem95}.
A detailed analysis with CEM95 of
more than 600 excitation functions for proton induced
reactions on 19 targets ranging from $^{12}$C to $^{197}$Au, for
incident energies ranging from 10~MeV to 5~GeV and a comparison to
available data, to calculations using approximately two dozen other
models, and to predictions of several phenomenological systematics
is presented in the
next paper by Mashnik et al.~\cite{mashnik97} in
this issue.
A comparison of many excitation functions calculated with CEM95 with
predictions of several other codes may be found in the recent
NEA/OECD document~\cite{Mich2}.
In present work, all calculations with CEM95 were
performed using for the macroscopic fission barriers
the Yukawa-plus-exponential modified LDM of Krappe,
Nix and Sierk~\cite{kns}, and Cameron's shell and pairing
corrections~\cite{cameron57} for the ground state and Barashenkov
et al.~\cite{barashenkov73} corrections for the saddle-point masses
for microscopic fission barriers;
the third Iljinov's et al.~systematics for the level
density parameters~\cite{iljinov92},
with shell corrections by Cameron et al.~\cite{cameron70},
without taking into account the dependence of fission barriers on
angular momenta,
with the dependence of fission barriers on excitation energy
proposed in Ref.~\cite{barashenkov74}, and with a
fixed value of 1.082 for the ratio $a_f/a_n$
for both incident energies. All values of other CEM parametrs are fixed
and are the same as in Refs.~\cite{cem,cem95}.\\

   The codes included into the CASCADE complex~\cite{cascade}
have been developed
and employed for many years in the Laboratory of Computing Technique and
Automation of JINR, Dubna to describe interactions of
hadron and nuclei with nuclei and with
gaseous and condensed matter.

Calculation of intranuclear cascades is performed in diffuse nucleon
clouds, the space distribution of which is defined from experiments on
electron scattering. The model takes into consideration the competition of
evaporation and fission processes in excited nuclei remaining after the
completion of the cascade and preequilibrium (described, according
to~\cite{modex}) processes.
   Inelastic interaction of the nuclei appears as a superposition of
two-particle nucleon-nucleon and pion-nucleon collisions which can
occur both in the nuclear overlap
region and inside each nucleus separately (see the second paper
in Ref.~\cite{cascade}).
Interaction of cascade particles with intranuclear
nucleons occur also in the case when the interacting nuclei are already
separated and continues till all of the cascade particles are either
absorbed by the nuclei or escape the nuclei. Mutual interactions of the
cascade particles are neglected.
   The number of nucleons in the clouds (Fermi particles) is equal to mass
numbers of striking nuclei in the beginning of the process and gradually
decreases due to knocking out of the nucleons by the growing shower of
cascading particles (the so-called ``trawling" effect). Therefore, the
coordinates of all nucleons are simulated in the beginning of the collision
and are redefined during the calculation procedure depending on the cascade
development. At a starting point $t$, corresponding to a moment when the
colliding nuclei come into contact, all particle collisions permitted by
kinematics and the Pauli principle are simulated. Only one which takes place
earlier than the others is chosen ($dt=min\{t_i\}$); after that the positions
of
the interacting nucleons and all cascading particles (nucleons and produced
pions) are shifted to the new points corresponding to the moment of time
$t+dt$. The calculation procedure is further repeated until all the
cascade particles
in the colliding nuclei are exhausted (see details in the third paper of
Ref.~\cite{cascade}).
   Decay of the excited nuclei is calculated basing on the evaporation model
as described in~\cite{book}
with a fixed level density parameter which is taken to be a=A/10 [MeV$^{-1}$].
   In order to consider the internuclear cascades in material samples the
codes are supplied with simulation of the particle transport including
energy losses on ionization, definition of the interaction point considering
the boundaries of the media with different chemical composition. Neutron
transport is calculated basing on the 26-group system of neutron
cross-sections.\\

The cascade -- preequilibrium -- evaporation -- fission code INUCL was
developed during several years at ITEP, Moscow by Stepanov with
co-authors~\cite{inucl}. The cascade stage of INUCL  was inspired by
the standard Dubna INC~\cite{book} but differs from it in several
points, as: use of new, more complete than in~\cite{book} experimental
data for cross sections of elementary NN and $\pi$N processes (37), the
range of incident particle energies was extended in INUCL
up to 15 GeV (in the standard
Dubna INC, it is only up to about 5 GeV), use of a different
radial distribution of
nuclear pairs absorbing pions inside a nucleus (38), taking into
account the local reduction of nuclear density (``trawling" effect)
at energies above several GeV
and several other details
(see the first paper of Ref.~\cite{inucl}).

The preequilibrium stage of INUCL was inspired by the MEM~\cite{mem,modex}
and CEM~\cite{cem} but differs from the Dubna version by using another
approach for the squared matrix elements of the transition rates and by
neglecting emission of complex particles and not
taking into account the forward peaked angular distributions
of preequilibrium
nucleons (see details in the second paper of Ref.~\cite{inucl}).

The evaporation model in INUCL was also inspired by the Dubna
version~\cite{book,modex}, except the use of other level density parameters
and, at low excitation
energies (below the separation
energy of nucleons or complex particles),  calculation of emission of
low-energy gammas (see details in the second paper of Ref.~\cite{inucl}).

The most interesting point of INUCL is incorporation at the evaporation
stage of reactions of a new, ``home-grown"
thermodynamical model of high energy fission (see the third and fourth
papers in Ref.~\cite{inucl}).
{}From a physical point of view,
the thermodynamical model of fission
is statistical by nature,
but provides
$A$-, $Z$- and energy-distributions of fission fragments quite different
from the well known Fong's
statistical model of nuclear fission~\cite{fong}, uses a
own systematics for the ratio of level density parameters
$a_f/a_n$, and seems to describe
experimental data better (see details in the third and fourth
papers of Ref.~\cite{inucl}).\\

The popular transport cascade -- evaporation code HETC~\cite{hetc}
was developed at Oak Ridge
National Laboratory. HETC is basically an extension of the code
NMTC~\cite{nmtc}  to allow particle transport up to several hundred
GeV~\cite{hetc}. For nonelastic collisions at energies below
$\sim 3$ GeV, HETC uses Bertini INC model~\cite{bertini} at the cascade
stages of reactions and Guthrie's evaporation program~\cite{hetcevap}
to determine the
energy and direction of emitted cascade nucleons and pions and evaporated
n, p, d, t, $^3$He and $\alpha$-particles, and the mass, charge,
and recoil energy of the residual nucleus.

As we mentioned above, a
detailed comparison of  the Bertini INC~\cite{bertini} with the
 Dubna INC~\cite{book} and
with Chen's et al. INC~\cite{vegas} may be found in
Ref.~\cite{jinrornlbnl}. We use on our work the standard version
of HETC~\cite{hetc}, well known and widely used in many
laboratories, therefore we do not consider necessary to repeat here
its detailed description.\\

LAHET is a Monte Carlo code for the transport and interaction of nucleons,
pions muons, light ions, and antinucleons in complex geometry~\cite{lahet};
it may also be used without particle transport to generate particle production
cross sections. LAHET is the result of a major effort at Los Alamos
National Laboratory to develop a code system based on the LANL version of the
HETC Monte Carlo code for the transport of nucleons,
pions and muons, which was originally developed at Oak Ridge National
Laboratory~\cite{hetc,hetc77}. Due to many new features added at LANL, the
code has been renamed LAHET, and the system of codes based on LAHET designated
as the LAHET Code System (LCS)~\cite{lahet}.

LAHET can use Bertini INC~\cite{bertini}
(from HETC) to describe nucleon-nucleus
interactions below 3.5 GeV, and a scaling law approximation to continue the
interaction energy to arbitrary high energies, although a reasonable upper
limit is about 10 GeV.

As an alternative to the Bertini INC, LAHET contains the INC routines from
ISABEL code. The ISABEL INC model is an extension by Yariv and
Fraenkel~\cite{isabel} of the VEGAS code~\cite{vegas} mentioned above.
ISABEL has the capability of treating nucleus-nucleus interactions as well
as particle-nucleus interactions. In the present work, we use both
Bertini INC and ISABEL versions of the INC to perform calculations with LAHET.

Let us mention here two more
points between the interesting features of LAHET (see details in~\cite{lahet}).
First, LAHET allows  one to calculate preequilibrium processes as an
intermediate stage between the intranuclear cascade and evaporation/fission
in the framework of a multistage multistep preequilibrium exciton model
(``MPM")~\cite{mpm}. As initially
suggested and used in the MEM~\cite{mem,modex},
the MPM uses the Monte Carlo method to solve the system of master equations
describing the process of equilibration of excited residual nucleus
remaining after the cascade stage of a reaction.
Nevertheless, there are several important differences between MPM and MEM.
First, the master equations of the MPM is simplified as compared to the
one of MEM: MPM takes into account only nuclear transitions
with $\Delta n = +2$, i.e., only in the direction of equilibration, while the
MEM considers all possible transitions
$\Delta n = +2, = -2$, and =0, accounting for all possible
positions of the
particle-hole pairs with respect to the Fermi level (transitions with
$\Delta n = 0$). Then, the master equations of the
MPM does not take into account the angular
distributions of preequilibrium particles
(it should be noted, that as an option, the MPM and
correspondingly the code LAHET, allow
to calculate angular distributions of
preequilibrium particles by using the phenomenological parametrization
of Kalbach (see~\cite{mpm})), while the MEM does this and the version of
the MEM used in CEM95
takes into account
the conservation of momentum and angular momentum of the nuclear system at
both preequilibrium and evaporation stages of a reaction. Then, the Monte
Carlo algorithms used in the MPM amd MEM for solving the corresponding master
equations also differ.
There are several other differences between the MPM and MEM as using
of different approximations for the inverse cross sections and
Coulumb barriers, for the level density parametrs and for the matrix elements
of
nuclear transitions, etc. (see details in~\cite{mem,modex,mpm}).
Note, that the
interface between the intranuclear cascade
model and the MPM and the interface between
MPM and the evaporation model~\cite{lahetevap} used in LAHET
and the ones used in CEM95
are also different.

Let us note also that LAHET includes as user options two models for
fission induced by high energy iteractions: the ORNL model~\cite{ornlfiss},
and the Rutherford Appelton Laboratory (RAL) model
by Atchinson~\cite{atchinson}; the fission models are employed with the
evaporation model. The RAL model allows fission for $Z \ge 71$, and
is the default fission model in LAHET. The RAL model really
is two models, for actinide and for subactinide fission. A detailed
description of the models incorporated in the LCS may by found in
Ref.~\cite{lahet} and is available on the World Wide Web in several
documents (see~\cite{lahet}).\\

The GNASH code~\cite{gnash}
applies the Hauser-Feshbach theory to calculate the
decay of compound nuclei in an open-ended sequence of decay chains,
conserving angular momentum and parity at all stages of the
reaction. Prior to equilibrium decay, the preequilibrium emission of
fast particles is accounted for using the exciton model, for up to two
preequilibrium ejectiles prior to equilibration. The exciton model also
includes angular momentum considerations to determine the spin
distributions of residual nuclei after preequilibrium emission~\cite{gnashp}.
Transmission coefficients are calculated from the optical model,
and the continuum (statistical) level density formalism of Ignatyuk et
al. is matched onto experimental low-lying discrete levels. In this
way, radionuclide production can be calculated to both the ground
state, and isomeric states.\\

The ALICE code calculates equilibrium decay with the Weisskopf-Ewing
theory, and preequilibrium emission is determined using the new Hybrid
Monte Carlo Simulation (HMS) model~\cite{alice96}. This model follows
successive interactions of excited nucleons creating three
quasi-particle excitations during the equilibration
process. Importantly, the theory is able to account for any number of
multiple preequilibrium processes in a natural way -- a major
advantage over some other preequilibrium models which only consider a
maximum of two preequilibrium ejectiles. The model has been applied
successfully at energies up to 400 MeV for the calculation of
excitation functions~\cite{alice96,alice962}.\\

Since most of simulation codes (except for GNASH) cannot
simulate the metastable states of product nuclei, the respective nuclide
chains used to simulate the cumulative yields of the product nuclei
 were taken to be simplified.

The results obtained by the codes are displayed in two sets of
figures, first, in Figs. 9 and 10, which show the product mass
distribution for $E_p$ = 130 MeV and 1500 MeV, respectively, and,
second, in Figs. 11 and 12, which present the
simulation-to-experiment ratios versus relative mass difference
between an initial target nucleus and a particular product nucleus
for $E_p$ = 130 MeV and 1500 MeV, respectively.
Together with the simulated results, Figs. 9 and 10 show experimental
data obtained by summing the yields with respect to all the available
isobar decay chains making allowance for their cumulating effect.

The $E_p$ = 1500 MeV experimental results shown in Fig. 10 are a visual
demonstration of
the different production channels, namely, due to spallation reactions for
masses ranging from $\sim$125 to 210 and due to fission reactions for
masses
ranging from $\sim$30 to $\sim$140. The range of masses from $\sim$125 to
$\sim$140
is most probably transient, where the reaction products can be generated by
both channels.
The $E_p$ = 130 MeV experimental results shown on Fig. 9 demonstrate only
the production
channel due to spallation reactions for masses ranging from $\sim$190 to
207.
The lack of experimental data for the fission product masses is accounted
for by the poor exposure conditions of the experimental samples.

Table 3 presents the experimental and simulated independent yields of
the ground states of $~^{209}$Bi(p,x)--reaction products at $E_p$ = 1500 MeV
and
130 MeV for some of the nuclear reaction product yields.
Table 4 presents the experimental and simulated values of the
independent and cumulative yields of the metastable states of
$~^{209}$Bi(p,x)--reaction
products at $E_p$ = 1500 MeV and 130 MeV. The simulation results
relate only to $E_p$ =  130 MeV since the GNASH code is valid for
simulation of proton interactions with initial energy below threshold
of pion productions ($\sim$ 0.2 GeV)

Table 5 presents the total experimental values of the
$~^{209}$Bi(p,x)--reaction
product ground and metastabe states at $E_p$ = 1500 MeV together
with
the respective simulated ground states.

\subsection*{9.3. Comparison between experiment and simulation}

Table 6 presents the information that shows the predictive power of each
code for both energies, namely, coincidence statistics which include
number of experimentally measured products $N_{EXP}$,
number of simulated products that were measured $N_S$,
number of ``coincidences" between simulated and experimental values
$N_C$  and the mean deviation $\langle H \rangle $ of simulation results
from
experimental data

Obviously, the simulation results shown on Figs. 9 and 10 do not contradict
the
experimental data if the simulated values run above the experiment and
repeat their general trend. The explanation is that the direct precision
$\gamma$-spectrometry technique used in this work makes it possible only to
identify the radioactive products of a high yield that includes a
significant part of the total yield of a given mass.
Unlike the cumulative yields, the independed yields makes possible to
performe the direct comparison with corresponding calculated values.

As seen from Tables 3, 4 and 5 the majority of the simulated results agree
well with the experimental values. However, in some cases  the simulation
results are underestimated significantly comparing
with the experimental data, especially for short-lived products.
There are a set of particular products whose simulated yields
described by the majority of the used codes are below the experimental
ones by several orders of magnitude.

The following conclusions concerning the scope of applicability of the
simulation codes can be drawn from comparing simulation and
experiment results:

\begin{itemize}
\item
INUCL describes quite reliably spallation products with
masses above A=180 and fission products with masses below A=120;
the products in the intermediate mass region A=120 to 180
are underestimated basically by a factor of 2 to 5; analysis
of
particular products shows that the code underestimates
significantly, up to a few orders of magnitude, the short-lived
products (see, e.g., the yields of $^{203}$Po and $^{202}$Po),

\item
HETC properly describes the spallation region, though, at $E_p = 150$
MeV, a too large underestimation for production of $^{201}$Pb and,
especially of $^{200}$Tl was obtained; since
the available version of the code does not include the fission
processes, it
cannot be used for predicting fission products,

\item
CEM95 also predicts only spallation products but does
it most perfectly compared with the other codes; a new
version of the CEM realized in the code CEM97 (now in progress,
see~\cite{mashnik97}) is expected to be even more
succesful in the spallation region and is extended to describe also
production of daughter nuclides in the fission and fragmentation regions,

\item
CASCADE describes satisfactorily the production of most nuclides in the
spallation region but overestimates some yields of nuclides in the
intermediate mass region from A=178 to about 195 and strongly underpredicts
production of several short-lived nuclides (see, e.g., the yields of
$^{200}$Tl, $^{198}$Tl in the spallation region and the yield of
$^{146}$Eu in the intermediate region),

\item
LAHET predicts the yields of most measured nuclides within a
factor of 2 both in the spallation and fission regions, as well as in the
intermediate mass region; nevertheless, some too
big discrepancies were obtained for $^{200}$Tl (at $E_p = 130$ MeV) in the
spallation region, for $^{146}$Eu and $^{140}$La in the intermediate
region, and for $^{82}$Br and $^{76}$As in the fission region,

\item
GNASH is the only code used here capable of calculating metastable states
and has predicted the yields of $^{202m}$Pb and $^{197m}$Pb well; it
also described within a factor of 2 the production of most ground state
isotopes in the spallation region, except the yields of $^{206}$Bi and
$^{201}$Pb,

\item
ALICE also predicted the yields of most isotopes in the spallation
region within a factor of 2, except production of $^{201}$Pb, $^{200}$Tl,
and $^{198}$Tl.
\end{itemize}

\section*{10. Conclusion}

This study is the first step in our work
on non-fissible targets of interest for accelerator-driven facilities.
Final conclusions about the predictive
properties of the codes should thus be drawn in a following paper.
At the present time,
we can draw only a preliminary conclusion that theoretical
yields predicted by different codes differ sometimes up to two orders of
magnitude. This is a strong indication
that further development of all codes is necessary before they can become
reliable predictive tools.

\section*{Acknowledgments}

We thank
N.V. Stepanov for useful
theory discussions and, especially, to F.K. Chukreev (Kurchatov
Institute) for his helpful comments concerning nuclear decay chains
data.
One of the authors (S.~G.~M.)
is grateful to R.~E.~MacFarlane, D.~G.~Madland,
P.~M\"oller, J.~R.~Nix, A.~J. Sierk, L.~Waters and P.~G.~Young of
LANL for many helpful discussions and support.

The work was made under the ISTC Project \# 017 and was completed under the
auspices of the U.S. Department of Energy by the Los Alamos National
Laboratory under contract no. W-7405-ENG-36.

\newpage

{\noindent
Table 1}\\
Approximation coefficients for the function
$\sigma=\sum_{i=0}^ka_iE^i, \qquad (\sigma\left[mb\right],
E\left[GeV\right])$  \\
\begin{center}

\begin{tabular}{ l l l l }
\hline
E     & 100-350 MeV   & 350-800 MeV   & 800-2600 MeV \\
      & k=4           & k=5&k=4                      \\
\hline
$a_0$ & .16082579E+2  & .91163994E+0  &.10809485E+2  \\
$a_1$ & -.98011691E+2 & .72710174E+2  &.12760849E+1  \\
$a_2$ & .48105521E+2  & -.22074848E+3 &-.25036816E+1 \\
$a_3$ & -.92886532E+2 & -25781415E+3  &.10439362E+1  \\
$a_4$ & .63086480E+3  & -.25781415E+3 &-.13817290E+0 \\
$a_5$ &               & .77078179E+2  &              \\
\hline
\end{tabular}
\end{center}
\vspace*{1.2cm}

\newpage

{\noindent
Table 2} \\
Experimental cumulative yields of the ground states of
$^{209}$Bi(p,x)--reaction
products at $E_p$= 1500 MeV and 130 MeV

\begin{center}

\begin{tabular}{ l l l l }
\hline
& & $E_p$=1500MeV & $E_p$=130MeV \\
\hline
Nucleus & Half life & Cross Section [mb] & Cross Section [mb] \\
\hline
${}^{*207}$Bi & 32.2 y   &  58.0 $\pm$ 7.5  & -\\
${}^{206}$Bi  & 6.243 d  &  26.2 $\pm$ 3.2  & 88.1 $\pm$ 9.3  \\
${}^{205}$Bi  & 15.31 d  &  24.5 $\pm$ 2.9  & 109  $\pm$ 23   \\
${}^{204}$Bi  & 11.22 h  &  25.2 $\pm$ 3.0  & 97.0 $\pm$ 11.4 \\
${}^{203}$Bi  & 11.73 h  &  19.4 $\pm$ 2.5  & 107  $\pm$ 13   \\
${}^{202}$Bi  & 1.67 h   &  11.0 $\pm$ 2.2  & 138  $\pm$ 15   \\
${}^{201}$Bi  & 108 m    &                  & 85.5 $\pm$ 11.7 \\
${}^{200}$Bi  & 36 m     &  8.4  $\pm$ 1.0  & 61.6 $\pm$ 8.0  \\
${}^{203}$Pb  & 51.837 h &  44.3 $\pm$ 4.9  & 230  $\pm$ 35   \\
${}^{201}$Pb  & 9.33 h   &  29.9 $\pm$ 5.0  & 139  $\pm$ 17   \\
${}^{200}$Pb  & 21.5 h   &  28.4 $\pm$ 3.3  & 122  $\pm$ 14   \\
${}^{199}$Pb  & 90 m     &  30.7 $\pm$ 5.8  & 113  $\pm$ 23   \\
${}^{198}$Pb  & 2.4 h    &  25.0 $\pm$ 5.0  & 20.0 $\pm$ 8.2  \\
${}^{201}$Tl  & 72.912 h &  38.7 $\pm$ 4.7  & 166  $\pm$ 27   \\
${}^{200}$Tl  & 26.1 h   &  36.2 $\pm$ 4.2  & 108  $\pm$ 13   \\
${}^{199}$Tl  & 7.42 h   &  32.9 $\pm$ 4.5  & 101  $\pm$ 18   \\
${}^{198}$Tl  & 5.3 h    &  27.4 $\pm$ 4.6  & 22.0 $\pm$ 3.7  \\
${}^{197}$Tl  & 2.84 h   &                  & 13.4 $\pm$ 2.4  \\
${}^{196}$Tl  & 1.84 h   &                  & 18.0 $\pm$ 5.3  \\
${}^{194}$Tl  & 33.0 m   &  11.4 $\pm$ 3.5  &                 \\
${}^{203}$Hg  & 46.612 d &  1.15 $\pm$ 0.12 &                 \\
${}^{195}$Hg  & 9.9 h    &  12.6 $\pm$ 3.8  &                 \\
${}^{192}$Hg  & 4.85 h   &  23.4 $\pm$ 3.6  &                 \\
${}^{198}$Au  & 2.696 d  &  0.46 $\pm$ 0.18 &                 \\
${}^{192}$Au  & 4.94 h   &  30.6 $\pm$ 4.9  &                 \\
${}^{191}$Pt  & 2.9 d    &  20.2 $\pm$ 4.1  &                 \\
${}^{188}$Pt  & 10.2 d   &  25.1 $\pm$ 2.9  &                 \\
${}^{188}$Ir  & 41.5 h   &  24.7 $\pm$ 2.8  &                 \\
${}^{186}$Ir  & 15.8 h   &  10.2 $\pm$ 1.8  &                 \\
${}^{184}$Ir  & 3.02 h   &  16.4 $\pm$ 3.7  &                 \\
${}^{*185}$Os & 93.6 d   &  29.3 $\pm$ 3.6  &                 \\
${}^{182}$Os  & 22.1h    &  23.4 $\pm$ 2.9  &                 \\
${}^{183}$Re  & 70.0 d   &  17.3 $\pm$ 2.9  &                 \\
${}^{182}$Re  & 12.7 h   &  26.3 $\pm$ 2.9  &                 \\
${}^{181}$Re  & 19.9 h   &  20.5 $\pm$ 4.5  &                 \\
${}^{179}$Re  & 19.7 m   &  16.4 $\pm$ 2.0  &                 \\
${}^{176}$Ta  & 8.08 h   &  22.6 $\pm$ 5.4  &                 \\
\hline
\end{tabular}

\end{center}

\newpage
{\noindent
Table 2 (continued)}\\

\begin{center}

\begin{tabular}{ l l l l }
\hline
& & $E_p$=1500MeV & $E_p$=130MeV \\
\hline
Nucleus & Half life & Cross Section [mb] & Cross Section [mb] \\
\hline
${}^{174}$Ta  & 1.18 h   & 15.6  $\pm$ 2.1  &   \\
${}^{173}$Ta  & 3.14 h   & 16.8  $\pm$ 3.1  &   \\
${}^{172}$Ta  & 36.8 m   &  6.6  $\pm$ 1.1  &   \\
${}^{173}$Hf  & 23.6 h   & 25.6  $\pm$ 3.1  &   \\
${}^{*172}$Hf & 1.87 y   & 13.2  $\pm$ 1.5  &   \\
${}^{170}$Hf  & 16.01 h  & 18.5  $\pm$ 2.6  &   \\
${}^{*173}$Lu & 1.37 y   & 23.2  $\pm$ 4.4  &   \\
${}^{171}$Lu  & 8.24 d   & 22.4  $\pm$ 2.5  &   \\
${}^{169}$Lu  & 34.06 d  & 16.2  $\pm$ 2.7  &   \\
${}^{166}$Yb  &   56.7 h & 11.1  $\pm$ 1.3  &   \\
${}^{160}$Er  & 28.58 h  &  9.6  $\pm$ 1.1  &   \\
${}^{157}$Dy  & 8.14 h   & 15.1  $\pm$ 1.8  &   \\
${}^{153}$Tb  & 2.34 d   &  8.2  $\pm$ 1.1  &   \\
${}^{*153}$Gd & 241.6 d  &  2.7  $\pm$ 0.8  &   \\
${}^{151}$Tb  & 17.609 d &  7.9  $\pm$ 0.9  &   \\
${}^{149}$Gd  & 9.4 d    &  8.3  $\pm$ 0.9  &   \\
$ {}^{147}$Gd & 38.1 h   &  6.6  $\pm$ 0.9  &   \\
$ {}^{146}$Gd & 48.27 d  &  6.1  $\pm$ 0.7  &   \\
${}^{147}$Eu  & 24.0 d   &  9.5  $\pm$ 1.2  &   \\
${}^{146}$Eu  & 4.59 d   &  6.7  $\pm$ 0.7  &   \\
${}^{145}$Eu  & 5.93 h   &  3.8  $\pm$ 0.6  &   \\
${}^{*143}$Pm & 265 d    &  1.07 $\pm$ 0.35 &   \\
${}^{139}$Ce  & 137.66 d &  2.4  $\pm$ 0.4  &   \\
${}^{140}$La  & 40.280 h &  1.3  $\pm$ 0.2  &   \\
${}^{140}$Ba  & 12.746 h &  0.82 $\pm$ 0.11 &   \\
${}^{127}$Xe  & 36.4 d   &  2.0  $\pm$ 0.2  &   \\
${}^{121}$Te  & 17.78d   &  1.4  $\pm$ 0.3  &   \\
${}^{113}$Sn  & 115.09 d &  0.75 $\pm$ 0.12 &   \\
${}^{111}$In  & 2.83 d   &  1.13 $\pm$ 0.15 &   \\
${}^{105}$Rh  & 35.36 h  &  4.3  $\pm$ 0.5  &   \\
$ {}^{95}$Tc  & 20.0 h   &  2.4  $\pm$ 0.5  &   \\
${}^{ 99}$Mo  & 2.75 d   &  4.5  $\pm$ 0.6  &   \\
${}^{95}$Nb   & 35.02 d  &  6.2  $\pm$ 0.9  &   \\
${}^{95}$Zr   & 64.02 d  &  1.65 $\pm$ 0.21 &   \\
${}^{89}$Zr   & 78.4 h   &  2.7  $\pm$ 0.4  &   \\
${}^{88}$Zr   & 83.4 d   &  0.22 $\pm$ 0.03 &   \\
${}^{ *88}$Y  & 106.61 d &  2.7  $\pm$ 0.5  &   \\
${}^{ 87}$Y   & 80.3 h   &  4.4  $\pm$ 0.5  &   \\
${}^{85}$Sr   & 64.08 d  &  5.2  $\pm$ 0.8  &   \\
${}^{ 83}$Rb  & 86.2 d   &  4.5  $\pm$ 0.9  &   \\
\hline
\end{tabular}

\end{center}

\newpage

{\noindent
Table 3} \\
Experimental and simulated independent yields (in mb) of the ground states
of $~^{209}$Bi(p,x)--reaction products at $E_p$= 1500 MeV and 130 MeV

\begin{center}

\begin{small}
\begin{tabular}{l l l l l l l l l l}
\hline
\multicolumn{10}{c} {$E_p$=1500MeV} \\
\hline
Nucleus & Half life&Exp. cr. sec.&CEM95&LAHET&INUCL&HETC&CASCADE&ALICE&GNASH \\
\hline
$^{206}$Po & 8.8 d  & 3.3$\pm$0.5  &1.3  &4.3  & 1.2 & 4.5 & 2.3 &      &  \\
$^{202}$Tl & 12.23 d& 4.8$\pm$0.7  &4.8  &4.6  & 8.6 & 7.3 & 3.1 &      &  \\
$^{200}$Tl & 26.1 h & 7.8$\pm$1.7  &6.1  &5.5  & 12.1& 7.1 & 4.6 &      &  \\
$^{146}$Eu & 4.59 d & 0.59$\pm$0.12&0.41 &5.1  & 0.15&     & 0.07&      &  \\
$^{140}$La & 40.28 h& 0.45$\pm$0.17&     &0.08 &     &     &     &      &  \\
$^{*102}$Rh& 2.9 y  & 0.69$\pm$0.12&     &1.5  & 2.9 &     &     &      &  \\
$^{95}$Nb  & 35.02 d& 3.8$\pm$0.5  &     &1.2  & 5.8 &     &     &      &  \\
$^{76}$As  & 26.32 h& 3.7$\pm$0.6  &     &0.80 & 2.7 &     &     &      &  \\
$^{74}$As  & 17.77 d& 2.1$\pm$0.3  &     &1.7  & 2.5 &     &     &      &  \\

\hline

\multicolumn{10}{c} {$E_p$=130MeV} \\

\hline
$^{207}$Po & 5.80 h & 24.5$\pm$3.8 & 34.8& 38.2& 20.6& 36.2& 45.2& 35.8& 33.6
\\
$^{206}$Po & 8.8 d  & 44.6$\pm$11.9& 39.1& 47.2& 28.0& 38.2& 68.5& 58.2& 32.1
\\
$^{205}$Po & 1.66 h & 30.5$\pm$3.7 & 45.9& 43.7& 25.7& 39.5& 60.2& 57.8& 21.2
\\
$^{204}$Po & 3.53 h & 43.1$\pm$5.2 & 45.2& 56.3& 22.5& 47.2& 85.6& 72.8& 25.5
\\
$^{203}$Po & 36.7 m & 38.5$\pm$8.9 & 47.6& 52.1& 3.2 & 51.0& 66.8& 57.1& 24.6
\\
$^{202}$Po & 44.7m  & 50.8$\pm$7.0 & 37.5& 62.3& 0.10& 57.9& 86.3& 67.3& 36.3
\\
$^{206}$Bi & 6.243 d& 57.7$\pm$6.5 & 79.1& 71.4& 66.0& 50.4&122.5&113.4& 168.1
\\
$^{201}$Pb & 9.33 h & 46.0$\pm$11.9& 58.9& 28.7& 25.1& 4.7 & 18.1& 10.7& 10.6
\\
$^{200}$Tl & 26.1 h & 15.4$\pm$8.1 & 6.2 & 1.9 & 16.5& 0.07& 0.17& 0.46& \\
$^{198}$Tl & 5.3 h  &  4.0$\pm$3.0 & 2.5 & 1.4 & 18.8&     & 0.47& 0.96& \\
\hline
\end{tabular}
\end{small}

\end{center}

\newpage

{\noindent
Table 4} \\
Experimental and simulated independent and cumulative yields of metastable
states of $~^{209}$Bi(p,x)--reaction products at $E_p$=1500 MeV and
$E_p$=130 MeV

\begin{center}

\begin{tabular}{l l l l l}
\hline
\multicolumn{5}{c} {$E_p$=1500MeV} \\
\hline

Nucleus     & Half life  & Yield & Experimental    & Calculation\\
            &            &       & cross sec. [mb] & (GNASH) [mb] \\
\hline
$^{204m}$Pb & 67.2 m     & ind   & 4.4$\pm$1.3     &  \\
$^{202m}$Pb & 3.53 h     & ind   &14.0$\pm$2.1     &  \\
$^{197m}$Pb &44.6 m      & cum   & 8.8$\pm$2.5     &  \\
$^{196m}$Tl & 1.41 h     & ind   &15.9$\pm$3.2     &  \\
$^{194m}$Tl & 32.8 m     & ind   & 4.4$\pm$0.6     &  \\
$^{193m}$Hg & 11.8 h     & ind   & 9.9$\pm$1.4     &  \\
$^{183m}$Os &9.9 h       & cum   &11.8$\pm$1.8     &  \\
$^{150m1}$Tb& 3.48 h     & ind   & 4.4$\pm$1.0     &  \\
$^{101m}$Rh &4.34 d      & cum   & 1.8$\pm$0.2     &  \\
$^{90m}$Y   &  3.19 h    & ind   & 2.4$\pm$0.4     &  \\
$^{82m}$Rb  &  6.472 h   & ind   & 2.5$\pm$0.4     &  \\

\hline

\multicolumn{5}{c} {$E_p$=130MeV} \\

\hline
$^{202m}$Pb & 3.53 h   & ind & 10.8$\pm$2.3 & 11.4\\
$^{197m}$Pb & 44.6 m   & cum & 13.4$\pm$2.4 & 10.6\\
\hline
\end{tabular}
\end{center}

\vspace*{2cm}

{\noindent
Table 5} \\
Experimental total yields [mb] of the ground and metastable states
of $~^{209}$Bi(p,x)--reaction
products at $E_p$=1500 MeV

\begin{center}

\begin{tabular}{l l l l l l l l l}
\hline
Nucleus   &Half life&Yield &Exp. cr. sec.&CEM95&LAHET&INUCL& HETC&CASCADE\\
\hline
$^{203}$Pb& 51.837 h&g+m1+m2 &25.9$\pm$5.8 & 13.8& 11.2& 18.8& 17.0& 10.3\\
$^{196}$Au&  6.183 d&g+m1+m2 &0.80$\pm$0.13& 1.6 & 0.93&  2.4& 0.22& 1.2\\
$^{194}$Au& 38.02 h &g+m1+m2 &1.37$\pm$0.24& 2.8 &  1.5&  3.6& 0.30& 1.8\\
$^{96}$Tc &  4.28 d &g+m     &0.96$\pm$0.22&     &  1.0&  2.8&     & \\
$^{82}$Br &  35.3 h &g+m     & 3.2$\pm$0.4 &     & 0.40&  2.3&     & \\
$^{*60}$Co& 5.2714 y&g+m     & 1.7$\pm$0.4 &     &  1.2& 0.64&     & \\
\hline
\end{tabular}

\end{center}

\newpage

{\noindent
Table 6}\\
Statistics of simulation-to-experiment comparisons. \\

\begin{center}
\begin{tabular}{l l l l l l l}
\hline
\multicolumn{1}{c} {} &
\multicolumn{3}{c} {E$_p$=130MeV} & \multicolumn{3}{c} {E$_p$=1500MeV} \\
\multicolumn{1}{c} {Code} &
\multicolumn{3}{c} {N$_{EXP}^{g+m}$=30, N$_{EXP}^g$=28 $^{\star 1}$} &
\multicolumn{3}{c} {N$_{EXP}^{g+m}$=100, N$_{EXP}^g$=89 $^{\star 1}$} \\
\cline{2-7}
& letter in& $N_C/N_S$ & $\langle H \rangle$ &
 letter in& $N_C/N_S$ &  $\langle H \rangle$ \\
& Fig. 11 &&&Fig. 12 &&\\
\hline
              &   &                  &      &    &       &      \\
CEM95         & A & 23/28            & 1.73 & A  & 49/64 & 1.95 \\
              &   &                  &      &    &       &      \\
INUCL         & B & 13/28            & 11.1 & B  & 54/86 & 2.95 \\
              &   & 13/23$^{\star 2}$&2.59  &    &       &      \\
              &   &                  &      &    &       &      \\
CASCADE       & C & 22/28            & 2.89 & C  & 50/64 & 1.83 \\
              &   &                  &      &    &       &      \\
HETC          & D & 19/26            & 3.22 & D  & 41/62 & 2.64 \\
              &   &                  &      &    &       &      \\
LAHET(ISABEL) & E & 24/28            & 1.86 &    &       &      \\
LAHET(Bertini)& F & 23/28            & 1.85 & E  & 63/89 & 2.51 \\
              &   &                  &      &    &       &      \\
ALICE         & G & 23/28            & 2.43 &    &       &      \\
              &   &                  &      &    &       &      \\
GNASH         & H & 8/10             & 1.98 &    &       &      \\
\hline
\end{tabular}
\end{center}
$^{\star 1}$ N$_{EXP}^{g+m}$ means number of all measured products,
N$_{EXP}^g$ means number of ground state measured products\\
$^{\star 2}$ with exception of $^{200}$Bi, $^{201}$Bi,
$^{202}$Bi, $^{202}$Po, $^{203}$Po

\newpage
\begin{center}
{\large \bf Figures Captions} \\
\end{center}

{\noindent
Fig. 1.
}
Plots of the recommended values of the monitor reaction excitation
functions (A - $^{27}$Al(p,x)$^{24}$Na, B - $^{27}$Al(p,x)$^{22}$Na).\\

{\noindent
Fig. 2.
}
A scheme of the extraction of the 800-2600 MeV external proton
beam. \\

{\noindent
Fig. 3.
}
A scheme of the extraction of the 70-200 MeV external proton
beam.\\

{\noindent
Fig. 4.
}
Relative distribution of the track numbers characterizing the
background of secondaries near the extracted 1500 MeV and 130 MeV proton
beams.\\

{\noindent
Fig. 5.
}
The $\gamma$-spectra of the $~^{209}$Bi(p,x)--reaction products
measured at proton energies of 1500 MeV and 130 MeV (the ``cooling" times
are
20 hours and 3.5 hours, respectively).       \\

{\noindent
Fig. 6.
}
Loading characteristics of the spectrometer.\\

{\noindent
Fig. 7.
}
Characteristics of the spectrometer (A - effictiveness, B -
height factor).\\

{\noindent
Fig. 8.
}
Time depenedences of the $^{209}$Bi (p,x)--reaction product decays
at E$_p$=1500 MeV (A) and E$_p$=130 MeV (B).\\

{\noindent
Fig. 9.
}
Product mass
distribution of  the $^{209}$Bi(p,x)--reaction  for $E_p$ = 130 MeV.\\

{\noindent
Fig. 10.
}
Product mass
distribution of  the $^{209}$Bi(p,x)--reaction  for $E_p$ = 1500 MeV proton.\\

{\noindent
Fig. 11.
}
Simulation-to-experiment ratios versus relative mass difference
between an initial target nucleus and a particular product nucleus
for $E_p$ = 130 MeV.
Letters are explained in Table 6.\\

{\noindent
Fig. 12.
}
Simulation-to-experiment ratios versus relative mass difference
between an initial target nucleus and a particular product nucleus
for $E_p$ = 1500 MeV.
Letters are explained in Table 6.

\end{document}